# Thermodynamic properties of neutral particle in presence of Topological defects in Magnetic Cosmic String Background

H. Hassanabadi [1] and M. Hosseinpour [*1]

[1]Physics Department, Shahrood of University, Shahrood, Iran P. O. Box: 3619995161-316, Shahrood, Iran
[*]Corresponding author, Tel.:+98 232 4222522; fax: +98 273 3395270
Email: hosseinpour.mansoureh@gmail.com

**Abstract**

In this paper, we study the covariant form of the non-relativistic Schrodinger-Pauli equation in the space-time generated by a cosmic string and discuss the solutions of this equation in present of interaction between the magnetic dipole momentum and electromagnetic field. We study the influence of the topology on this system. We obtain the solution of radial part as well as the energy levels. We consider all thermodynamic properties of neutral particle in magnetic cosmic string background by using an approach based on the partition function method.

## I. Introduction

The interaction between electric and magnetic fields and multipole moments has attracted a great deal of studies, such as [1-10], the holonomic quantum computation [11,12] and the Landau quantization [13,14]. Besides, recent studies of the interaction between a moving electric moment and external fields [8,9] have shown a difference between the field configuration that yields the arising of geometric phases for an electric charge [1,2,15], an electric dipole moment [3,7] and a moving electric moment. The quantum mechanics dynamics on conical spaces in the presence of topological defects has attracted much attention in recent years [16-18]. Among different motivations we can recall the context of the (2 + 1) dimensional quantum gravity [19], cosmic strings [20].The simple, but nontrivial, geometry of the cone appears as an effective geometry in such diverse physical entities as cosmic strings [21]. Accordingly, the dynamics of quantum neutral particles in a conical background has been studied with very different motivations [22-25]. An important issue concerning the cone is the fact that the conical background is naturally associated to a curvature singularity at the cone tip. The quantum dynamics of a single particle in a conical space-time has been investigated by several authors. Till now, some problems have been investigated in conical space-time including classical and quantum non-relativistic dynamics of a particle [26] and the influence of conical singularities in the energy levels of a harmonic oscillator [27].

In our work, we are going to discuss on the solution of non-relativistic Schrodinger-Pauli equation produced by a cosmic string. The latter is a linear defect that changes the topology of the medium. Topological defects will arise in some of the models with spontaneous symmetry breakdown in field theory. The cosmic strings are expected to have large mass density and very thin width. The space-time geometry around an infinitely stretching straight string has a peculiar property. From the field theory point of view, the cosmic string can be viewed as a consequence of symmetry breaking phase transition in the early universe [28]. Till now, some problems have been investigated in curved space–time including the one-electron atom

This paper is organized as follows. In Section II, we first review the covariant non-relativistic Schrodinger-Pauli equation in the space-time generated by a cosmic string. We next report the solution of radial part as well as the energy levels. Finally, in section III, We consider all thermodynamic properties of neutral particle in magnetic cosmic string background by using an approach based on the partition function method.

## II. Neutral particle in Cosmic String Background

The cosmic string space-time with an internal magnetic field in cylindrical coordinates is described by the line element [29-31].

$$ds^2 = -dt^2 + dr^2 + \alpha^2 r^2 d\varphi^2 + dz^2 \tag{1}$$

with $-\infty < z < \infty$, $r \geq 0$ and $0 \leq \varphi \leq 2\pi$. The parameter α is related to the linear mass density $\tilde{m}$ of the string via $\alpha = 1 - 4\tilde{m}$ and varies in the interval $(0,1]$

We can build the local reference frame through a non-coordinate basis with $e_\mu^{(a)}$ where $e_\mu^{(a)}$ and $e_{(a)}^\mu(x)$ are transformation matrices. The components of the non-coordinate basis $e_\mu^{(a)}$ are called tetrads or vierbeins that form our local reference frame and $e_{(a)}^\mu(x)$ satisfy

$$\eta^{ab} e_{\bar{a}}^\mu(x) e_{\bar{b}}^\nu(x) = g^{\mu\nu}(x) \tag{2}$$

Where $\mu, \nu = 0,1,2,3$ are tensor indices and $\bar{a}, \bar{b} = 0,1,2,3$ denote tetrad indices [32,33]. We can obtain the component of spin connection from

$$\omega_{\mu\bar{b}}^{\bar{a}} = e_\nu^{\bar{a}} e_{\bar{b}}^\sigma \Gamma_{\sigma\mu}^\nu + e_\nu^{\bar{a}} \partial_\mu e_{\bar{b}}^\nu \tag{3}$$

$\Gamma_{\mu\nu}^\sigma$ are The Christoffel symbols of the second kind. The non-vanishing components of the spin connection are

$$\omega_\varphi^{\overline{12}} = e_\mu^{\bar{1}} e^{\nu\bar{2}} \Gamma_{\varphi\nu}^\mu - e^{\nu\bar{2}} \partial_\varphi e_\mu^{\bar{1}} = = 1 - \alpha$$
$$\omega_\varphi^{\overline{12}} = -\omega_\varphi^{\overline{21}} \tag{4}$$

The relativistic dynamics of the neutral particle in this curved spacetime was studied in [34]. In the same paper, the non-relativistic behavior of the neutral particle in curved spacetime was obtained through the application of the Foldy-Wouthuysen approximation [35] to the Dirac equation. We assume that the dipole magnetic moments are parallel to the z-axis of the spacetime. The non-relativistic equation is [10]

$$i\frac{\partial \psi}{\partial t} = m\psi + \left[\frac{1}{2m}\vec{\pi}^2 - \frac{\mu^2 E^2}{2m} + \frac{\mu}{2m}\vec{\nabla}.\vec{E} + \mu\hat{n}.\vec{B}\right]\psi \qquad (5)$$

The unit vector $\hat{n}$ indicates the direction of the magnetic dipole moment. We introduce the generalized momentum momentum in presence of electromagnetic field as

$$\vec{\pi} = -i\hbar\vec{\nabla} + \left(\mu(\hat{n}\times\vec{E}) + \frac{(1-\alpha)}{2}\hat{\varphi}\right)_j \qquad (6)$$

Where

$$\vec{\nabla} = \frac{\partial}{\partial r}\hat{r} + \frac{1}{\alpha r}\frac{\partial}{\partial \varphi}\hat{\varphi} + \frac{\partial}{\partial z}\hat{z} \qquad (7)$$

We choose that the electric field as

$$\vec{E} = \frac{\lambda\rho}{2}\hat{\rho} \qquad (8)$$

With λ is a linear density charge. By substituting in Eq. (5) and by using the $\Psi = e^{-iEt}\psi$ we have

$$-\frac{1}{2m}\left(\frac{\partial^2\psi}{\partial\rho^2} + \frac{1}{\rho}\frac{\partial\psi}{\partial\rho} + \frac{1}{\alpha^2\rho^2}\frac{\partial^2\psi}{\partial\varphi^2} + \frac{\partial^2\psi}{\partial z^2}\right) - \frac{i}{2m}\left(\frac{\mu\lambda}{\alpha} + \frac{(1-\alpha)}{\alpha^2\rho^2}\right)\frac{\partial\psi}{\partial\varphi} + \frac{\mu^2\lambda^2}{8m}\rho^2\psi + \frac{1}{8m}\frac{\left((1-\alpha)^2\right)}{\alpha^2\rho^2}\psi$$

$$\frac{\mu\lambda}{2m}\psi + \frac{\mu\lambda}{4m}\frac{(1-\alpha)}{\alpha}\psi = E\psi \qquad (9)$$

The solution of the Schrodinger-Pauli equation can be written in the form[10]

$$\psi_{nl}(\rho,\varphi,z) = e^{il\varphi}e^{ikz}R_{nl}(\rho) \qquad (10)$$

By substituting in Eq(9)

$$\left[-\frac{1}{2m}\left(\partial_\rho^2 + \frac{1}{\rho}\partial_\rho\right) + \frac{k^2}{2m} + \frac{\gamma^2}{2m\alpha^2\rho^2} + \frac{\gamma\omega}{2\alpha} + \frac{m\omega^2}{8}\rho^2 + \frac{\omega}{2}\right]R_{nl}(\rho) = ER_{nl}(\rho) \qquad (11)$$

Where we defined $\gamma = l + \frac{(1-\alpha)}{2}$ and $\omega = \mu\lambda/m$. we make a convenient change of variables as

$$\xi = \frac{m\omega}{2}\rho^2$$

Then we have[10]

$$R''_{nl} + \frac{1}{\xi}R'_{nl} + \frac{1}{\xi^2}\left(\xi\beta - \frac{\gamma^2}{4\alpha^2} - \frac{\xi^2}{4}\right)R_{nl} = 0 \tag{12}$$

Where

$$\beta = \frac{E}{\omega} - \frac{k^2}{2m\omega} - \frac{\gamma}{2\alpha} - \frac{1}{2} \tag{13}$$

We can solve this equation. The corresponding wave functions and energy eigenvalues are obtained as

$$\psi_{nl} = N\,\xi^{\frac{1}{2}+\sqrt{\frac{1}{4}+\frac{\gamma^2}{4\eta^2}}}\, e^{-\frac{\xi}{2}} L_n^{1+2\sqrt{\frac{1}{4}+\frac{\gamma^2}{4\eta^2}}}(\xi) \tag{14}$$

$$\varepsilon = \left(n + \frac{a}{2}\right)\omega \tag{15}$$

Where

$$a = \left(2 + \sqrt{1 + \frac{\gamma^2}{\alpha^2}} + \frac{k^2}{m\omega} + \frac{\gamma}{\alpha}\right) \tag{16}$$

And N is the normalization constant.

## III.  Thermodynamic properties of system

In order to consider thermodynamic properties of neutral particle in magnetic cosmic string background for a constant $\ell$ we concentrate, at first, on the calculation of the partition function

$$Q_1 = \sum_{n=0}^{\infty} e^{-\beta\left(n+\frac{a}{2}\right)\omega} = e^{-\beta\omega\frac{(a-1)}{2}}\left\{2\sinh\frac{\beta\omega}{2}\right\}^{-1} \tag{17}$$

Where $\beta = 1/\kappa T$. The partition function for N-body system with no interaction inside obtain via

$$Q_N = (Q_1)^N = e^{-N\beta\omega\frac{a}{2}}\left\{e^{\beta\omega}-1\right\}^{-N} \tag{18}$$

Once the Helmholtz free energy is obtained, the other statistical quantities are obtained in a straightforward manner as

$$A = -\frac{1}{\beta}\ln Q_N = N\omega\frac{(a-1)}{2} + NKT\,\ln\left(2\sinh\frac{\beta\omega}{2}\right) \tag{19}$$

Chemical potential can obtain as

$$\mu = \frac{\partial A}{\partial N} = \omega \frac{(a-1)}{2} + KT \ln\left(2\sinh\frac{\beta\omega}{2}\right) \tag{20}$$

and the pressure is zero

$$P = -\frac{\partial A}{\partial V} = 0 \tag{21}$$

Once the Helmholtz free energy is obtained, the other statistical quantities are obtained in a straightforward manner. The mean energy is

$$U = -\frac{\partial \ln Q_N}{\partial \beta} = N\omega\left(\frac{a-1}{2} + \coth\frac{\beta\omega}{2}\right) \tag{22}$$

The main statistical quantity, i.e., the entropy, is related to other quantities via

$$\frac{S}{K} = \beta^2 \frac{\partial A}{\partial \beta} = \left(-N \ln\left(2\sinh\frac{\beta\omega}{2}\right) + N\beta\frac{\omega}{2}\coth\frac{\beta\omega}{2}\right) \tag{23}$$

the specific heat capacity at constant volume is obtained from

$$\frac{C}{K} = -\beta^2 \frac{\partial U}{\partial \beta} = -\beta^2 \omega^2 N \frac{e^{-\beta\omega}}{1-e^{-\beta\omega}} \tag{24}$$

One can verify that in the limit $\alpha \to 1$, the spacetime becomes flat. we recover the general solution for flat space-time. in this limit Eq. (12) rewritten as

$$R''_{nl} + \frac{1}{\xi}R'_{nl} + \frac{1}{\xi^2}\left(\xi\beta - \frac{\gamma^2}{4} - \frac{\xi^2}{4}\right)R_{nl} = 0 \tag{25}$$

In this case $\gamma = l$ and

$$\beta = \frac{E}{\omega} - \frac{k^2}{2m\omega} - \frac{l}{2} - \frac{1}{2} \tag{26}$$

The corresponding wave functions and energy eigenvalues are obtained from NU method as

$$\psi = N'\xi^{\frac{1}{2}(1+\sqrt{1+l^2})} e^{-\frac{\xi}{2}} L_n^{1+\sqrt{1+l^2}}(\xi) \tag{27}$$

$$\varepsilon = \left(n + \frac{a'}{2}\right)\omega \tag{28}$$

Where

$$a' = \left(2 + \sqrt{1+l^2} + \frac{k^2}{m\omega} + l\right) \tag{29}$$

and $N'$ is the normalization constant. In limit $\alpha \to 1$ the thermodynamic properties of system will be obtained from Eqs. (17-24) the parameter $a$ must be replaced by $a'$

In what following we depict the thermodynamic properties of the system vs. $KT$ in Figs. (1–6). In Fig. (1), where the Helmholtz free energy is plotted vs. $KT$ for different angular quantum number. In the interval $0 \leq KT \leq 100$, it is seen that the energy decreases in a nearly linear behavior for increasing $KT$. In Fig. (2), the thermodynamic energy is plotted vs. $KT$ for different angular quantum number in the interval $0 \leq KT \leq 100$. It reveals that for increasing $KT$, the internal energy is linearly increased as well. As we see in Figs.(1,2) the Helmholtz free energy and thermodynamic energy both are increase by increasing angular quantum number. Fig. (3), where the Helmholtz free energy is plotted vs. $KT$ for different $\alpha$. In the interval $0 \leq KT \leq 100$, the energy decreases in a nearly linear behavior for increasing $KT$. Fig. (4), represents the energy behavior vs. $KT$ for different $\alpha$ in the interval $0 \leq KT \leq 100$. as expected the thermodynamic properties tend to their behavior in flat space time when $\alpha$ tend to 1. Fig. (5) reveals that the entropy is slowly increasing for large $KT$ values. The curve is repeated for $\omega = 1, 2, 3, 4$. The variation of heat capacity vs. $KT$ for various $\omega$ values is shown in Fig. (6). As can be seen in Figs.(5,6), the entropy and the heat capacity decreases for increasing $\omega$. As we see in Figs. (7, 8) the entropy and the heat capacity are independent of $\alpha$. As expected in fig.(8) for increasing $KT$, the heat capacity tend to its saturation value in high temperature.

**Conclusion**

With the purpose of discussing the role of the topology on the Landau quantization we obtain the thermal properties of neutral particles with a permanent magnetic dipole moment interacting with an external magnetic field. In this way, some new results to the interesting problem considered in seminal papers by Bakke [10] about the effects of gravitational fields on breaking the infinite degeneracy of the Landau-Aharonov-Casher levels obtained in flat space-time. The presence of a cosmic string changes the solution as compared with the flat Minkowski spacetime results that are due to the combined effects of the curvature and the nontrivial topology determined by the deficit solid angle associated with this spacetime. We have investigated the influence of the topological defects on thermal properties. We obtain the solution of radial part as well as the energy levels. We consider all thermodynamic properties of neutral particle in magnetic cosmic string background by using an approach based on the partition function method. We have also analyzed the thermodynamic properties behavior graphically. We obtain that, by increasing the parameters of the topology $\alpha$, the entropy, heat capacity and free energy will also increase. the entropy and the heat capacity are independent of parameters of the topology. As we expect when $\alpha \to 1$, we recover the general solution for flat space-time.

**Acknowledgments**

It is a pleasure for authors to thank the kind referees for their many useful comments on the manuscript.

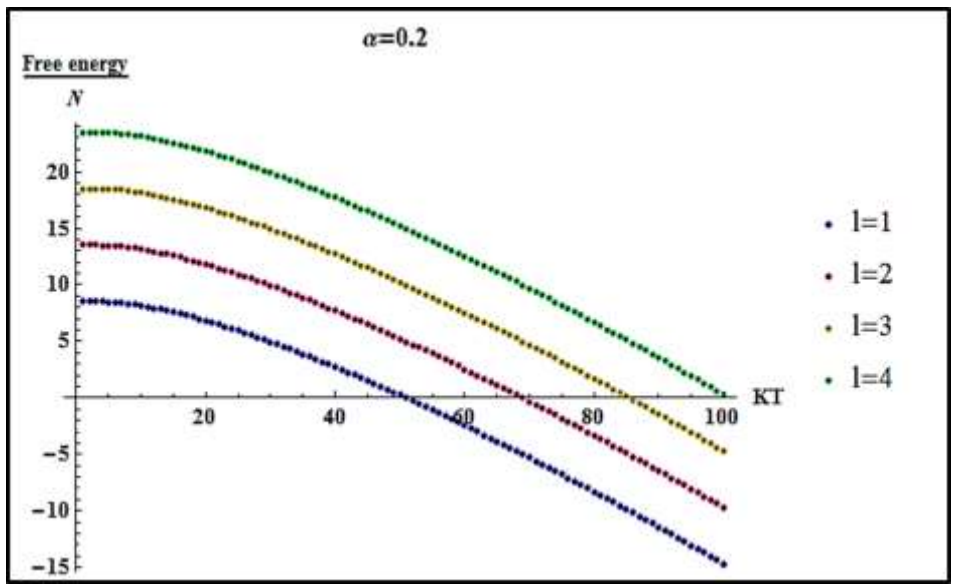

Fig(1). The comparison of the Helmholtz free energy as a function of $KT$ for different angular quantum number.

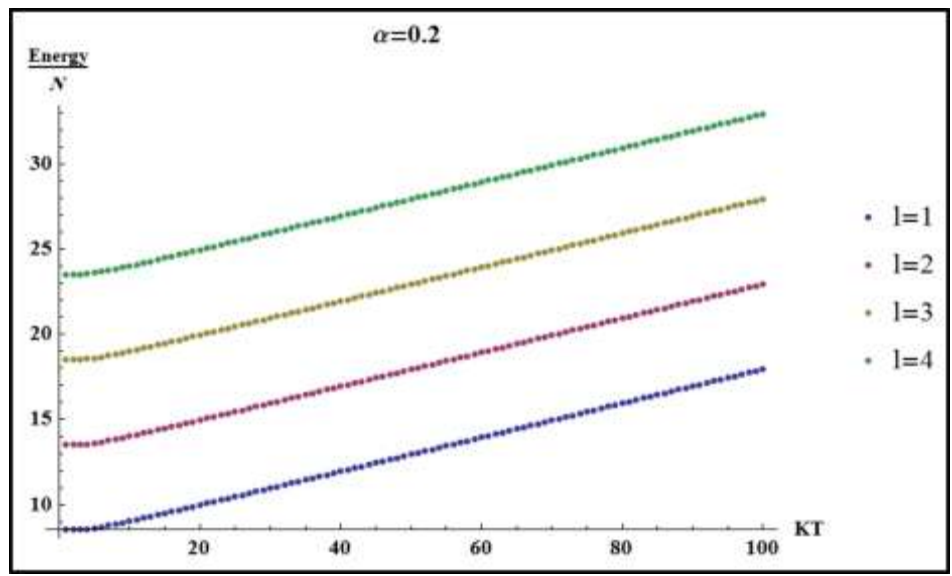

Fig.(2). The comparison of the thermodynamic energy $U/N$ as a function of $KT$ for different angular quantum number.

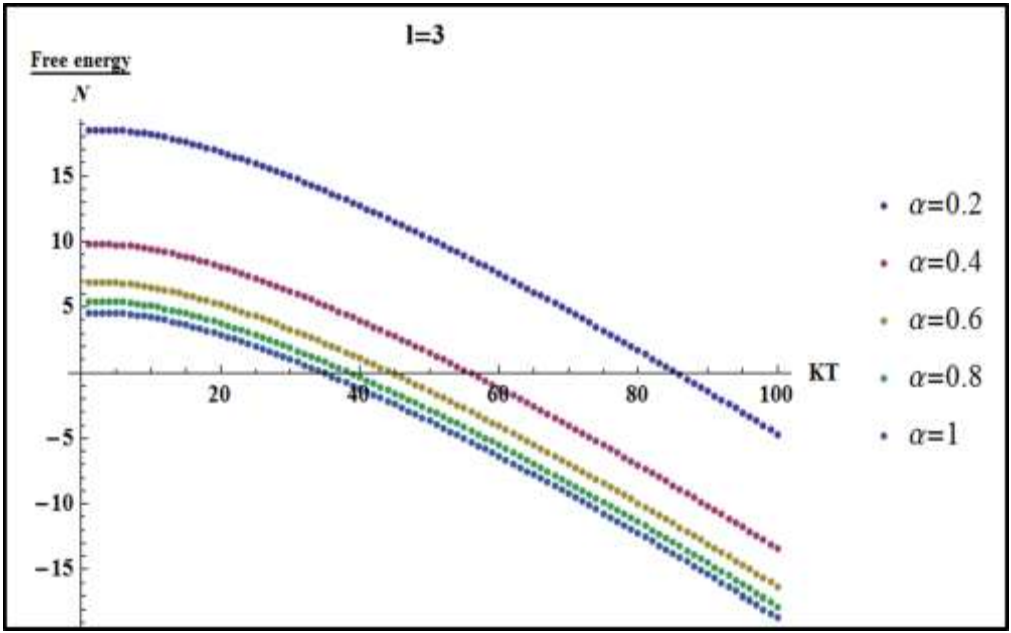

Fig.(3). The comparison of the Helmholtz free energy as a function of $KT$ for different $\alpha$

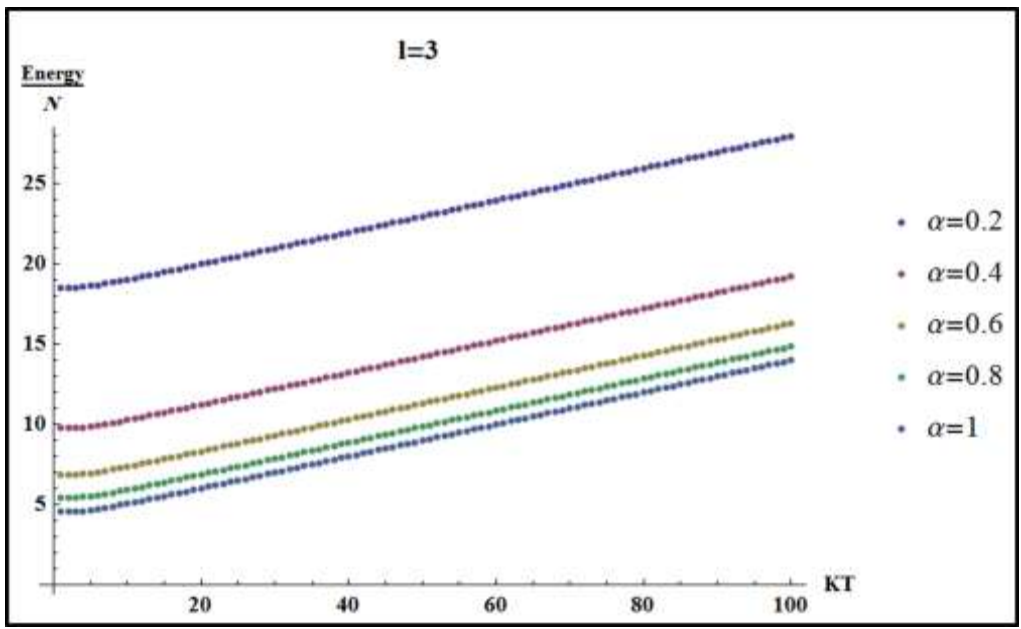

Fig.(4). The comparison of the thermodynamic energy $U/N$ as a function of $KT$ for different $\alpha$

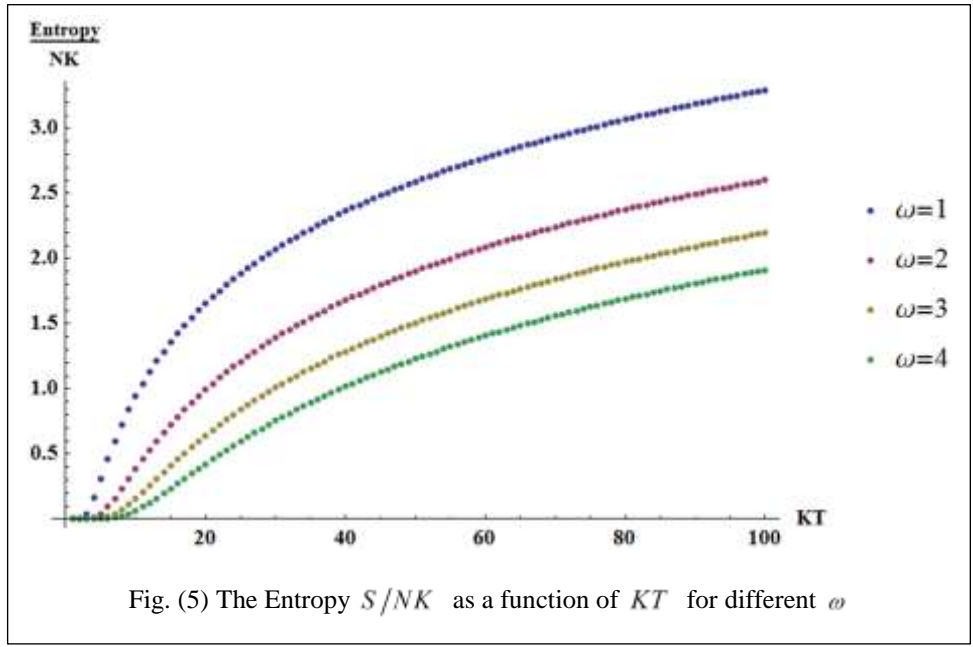

Fig. (5) The Entropy $S/NK$ as a function of $KT$ for different $\omega$

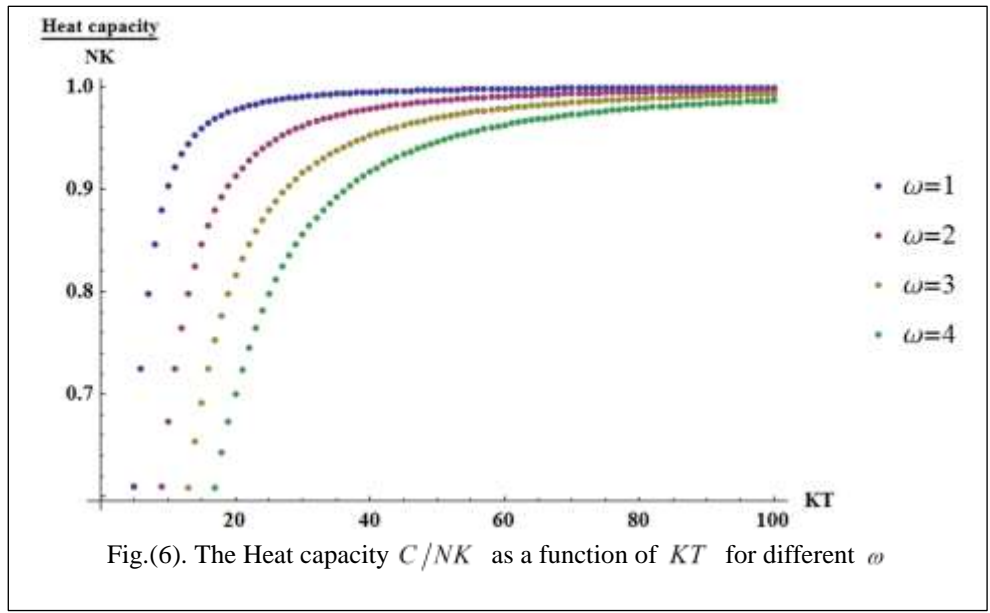

Fig.(6). The Heat capacity $C/NK$ as a function of $KT$ for different $\omega$

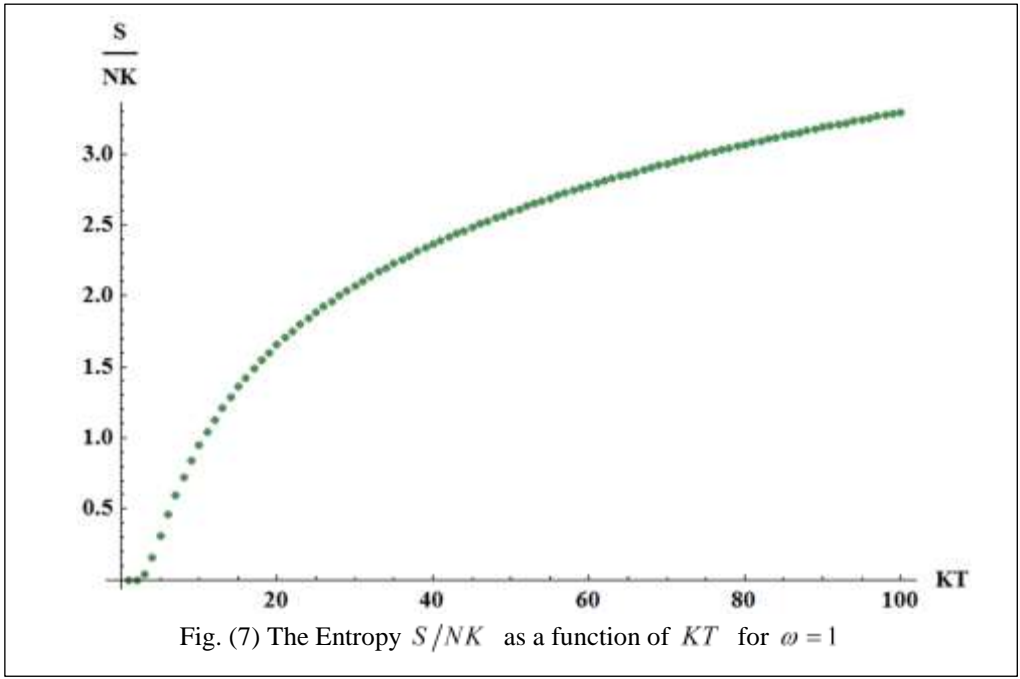

Fig. (7) The Entropy $S/NK$ as a function of $KT$ for $\omega = 1$

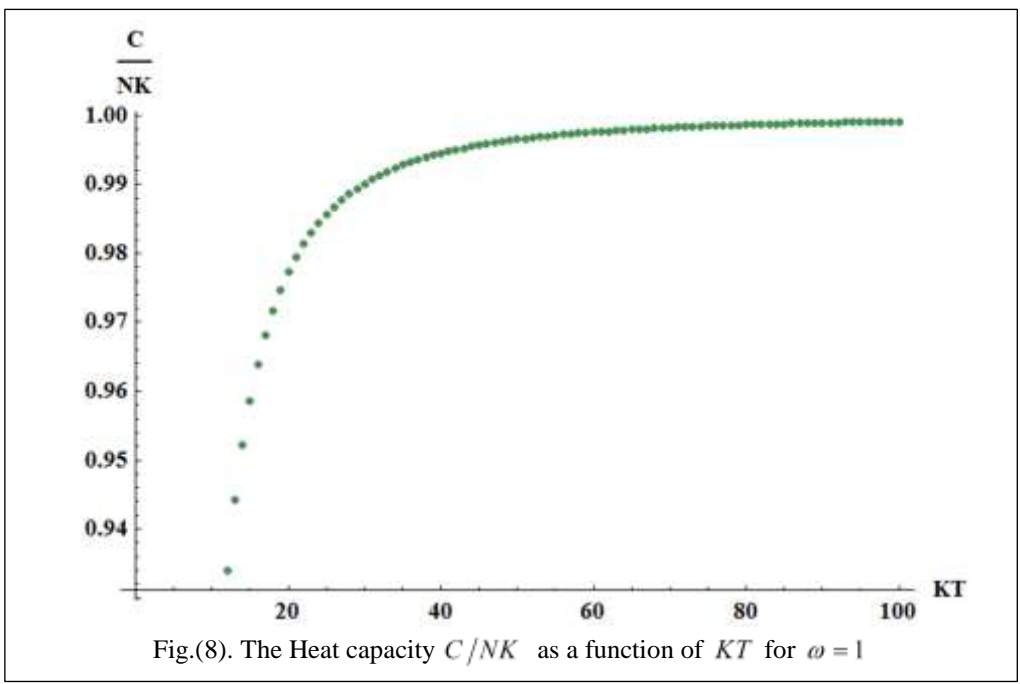

Fig.(8). The Heat capacity $C/NK$ as a function of $KT$ for $\omega = 1$